\documentclass[]{aa}

\usepackage{amsmath}
\usepackage{txfonts}
\usepackage{booktabs}
\usepackage{array}
\usepackage{graphicx}
\usepackage{natbib,twoopt}
\usepackage{upgreek}
\usepackage{color}
\usepackage{xcolor}
\usepackage{wasysym}
\usepackage[normalem]{ulem}
\usepackage[colorlinks,linkcolor=blue,anchorcolor=blue,citecolor=blue,breaklinks=true]{hyperref}
\usepackage{tablefootnote}
\usepackage{gensymb}
\bibpunct{(}{)}{;}{a}{}{,} 
\usepackage{placeins}

\newcommand{\hii}{H\,{\sc ii} }

\begin{document}

\title{The low-frequency flattening of the radio spectrum of giant \hii regions in \object{M\,101}}
\author{L. Gajovi\'c\inst{1} \and 
V. Heesen\inst{1} \and
M. Br\"uggen\inst{1} \and
H. W. Edler\inst{2,1} \and
B. Adebahr\inst{3,2} \and 
T. Pasini\inst{4} \and
F. de Gasperin\inst{4} \and
A. Basu\inst{5,6} \and
M. Weżgowiec\inst{7} \and
C. Horellou\inst{8} \and
D.J. Bomans\inst{2} \and
H. Dénes\inst{9} \and
D. Vohl\inst{10,2}
}

\institute{Hamburger Sternwarte, University of Hamburg, Gojenbergsweg 112, 21029 Hamburg, Germany
\and ASTRON, Netherlands Institute for Radio Astronomy, Oude
Hoogeveensedijk 4, Dwingeloo, 7991 PD, The Netherlands
\and Ruhr University Bochum, Faculty of Physics and Astronomy, Astronomical Institute (AIRUB), Universit\"atsstrasse 150, 44801 Bochum, Germany 
\and INAF - Istituto di Radioastronomia, via P. Gobetti 101, 40129 Bologna, Italy
\and Th\"uringer Landessternwarte, Sternwarte 5, 07778 Tautenburg, Germany
\and Max-Planck-Institut f\"ur Radioastronomie, Auf dem H\"ugel 69, 53121 Bonn, Germany
\and Obserwatorium Astronomiczne Uniwersytetu Jagiello\'nskiego, ul. Orla 171, 30-244 Krak\'ow, Poland 
\and Chalmers University of Technology, Onsala Space Observatory, SE-43992 Onsala, Sweden
\and School of Physical Sciences and Nanotechnology, Yachay Tech University, Hacienda San José S/N, 100119, Urcuquí, Ecuador
\and Anton Pannekoek Institute for Astronomy at the University of Amsterdam, University of Amsterdam, P.O. Box 94249, 1090 GE Amsterdam, The Nederlands
}

\date{Received 15 November 2024 / Accepted 6 February 2025}

\abstract
{ In galaxies, the flattening of the spectrum at low radio frequencies below 300 MHz has been the subject of some debate. A turnover at low frequencies could be caused by multiple physical processes, which can yield new insights into the properties of the ionised gas in the interstellar medium.}
{We investigate the existence and nature of the low-frequency turnover in the \hii regions of M\,101.}
{We study the nearby galaxy M\,101 using the LOw Frequency ARray (LOFAR) at frequencies of 54 and 144\,MHz, Apertif at 1370\,MHz, and published combined map from the Very Large Array (VLA) and Effelesberg telescope at 4850\,MHz. }
{The spectral index between 54 and 144\,MHz is inverted at the centres of \hii regions. We find a significant low-frequency flattening at the centres of five out of six \hii regions that we selected for this study. }
{The low frequency flattening in \hii regions of M\,101 can be explained with two different free-free absorption models. The flattening is localised in a region smaller than 1.5\,kpc and can only be detected with high resolution (better than 45\arcsec). The detection of low frequency flattening has important consequences for using radio continuum observations below 100\,MHz to measure extinction-free star-formation rates.}


\maketitle

\section{Introduction}

Non-thermal (synchrotron) and thermal (free-free) radiation processes play a key role in the radio continuum emission of nearby galaxies. Synchrotron radiation in galaxies is produced by cosmic-ray (CR) electrons that spiral around magnetic field lines and free--free radiation by free electrons scattering off ions \citep[e.g.][]{Niklas_1997, Basu_2012, Tabatabaei_2017}. The synchrotron radiation dominates the spectrum at radio frequencies below a few GHz. CRs that produce synchrotron radiation originate from regions with higher star formation in spiral galaxies because they are accelerated through shocks during supernova explosions \citep{Green_2014, Caprioli_2010}.
The spectrum of synchrotron radiation is a power law $S_\nu \propto \nu^\alpha$, with a radio spectral index $\alpha$ between $-1.2$ and $-0.5$ for nearby galaxies \citep[e.g.][]{Gioia_1982,Li_2016,Tabatabaei_2017}. At higher frequencies (around 10 GHz), the contribution of thermal free-free emission from ionised gas with the spectral index $\alpha=-0.1$ becomes significant and the spectrum should flatten out \citep[e.g.][]{Tabatabaei_2017, Klein_2018}. Lower-frequency (below 100 MHz) observations at high spatial resolution only recently became possible with the advent of digital interferometers, such as the LOw Frequency ARray \citep[LOFAR;][]{vanHarleem_2013}, and sophisticated calibration strategies \citep{vanWeeren_2016, Tasse_2018,deGasperin_2019}. Still, the spectral behaviour in this frequency regime remains largely unexplored, but there is evidence that the spectrum of nearby galaxies flattens or even turns over below 100 MHz.

The low frequency flattening was first noticed in statistical studies of galaxy samples. \citet{Israel_1990} observed that the radio intensities at 57.5 MHz are systematically lower than the values extrapolated from higher frequency measurements assuming a power law spectrum. They reported a dependence of the flattening on the inclination of the galaxies and attributed it to free-free absorption. However, \citet{Hummel_1991} was unable to find any correlation using the same data and stressed that a steepening of the spectrum at high frequencies due to propagation effects of the relativistic electrons is equally plausible. A more recent study by \citet{Chyzy_2018}, using additional data at 150\,MHz, confirms the results of \citet{Hummel_1991}. 

An alternative explanation for the low frequency flattening was provided by \citet{Pohl_1990} and \citet{Pohl_1991a,Pohl_1991b} who modelled the energy spectra of CR electrons that experience simultaneous injection, diffusion, convection, adiabatic deceleration, as well as radiative and ionisation losses. Observing a large sample of 250 bright galaxies at three pairs of frequencies with the lowest being 74 and 325 MHz, \citet{Marvil_2015} reports a curvature in the mean spectrum with a magnitude of approximately $\Delta\alpha=-0.2$ per logarithmic frequency decade. They propose two reasons for this effect: free-free absorption at low frequencies or a curved synchrotron spectrum resulting from curvature in the underlying energy distribution of CR electrons. 

To observe the low-frequency spectral flattening in individual galaxies it is usually necessary to look at exceptional targets where the flattening is significant at comparatively high frequencies. In the centre of the Milky Way, thermal absorption shapes the spectra below 500\,MHz \citep{Roy_2006}. Meanwhile, the integrated spectrum of the Milky Way has a detectable turnover at about 3\,MHz \citep{Brown_1973}. In the starburst galaxy M\,82, turnovers were detected in the global integrated radio spectrum \citep{Klein_1988}, the core regions \citep{Adebahr_2013}, and individual sources near the galactic centre \citep{Adebahr_2017}. There is also a low frequency turnover in the integrated spectrum of the starburst galaxy NGC\,253 \citep{Marvil_2015}. The ultra-luminous infrared galaxy Arp\,220 shows a turnover in the central region at around 1\,GHz \citep{Varenius_2016} and in the integrated spectrum \citep{Anantharamaiah_2000}. 

More recently, \cite{2022A&A...658A...4R} investigate the luminous infrared galaxy Arp 299 using sub-arcsecond LOFAR observations at 150 MHz. In order to fit the SED in the nucleus of this galaxy, they use two models for the free-free absorbing and emitting thermal gas. In the first model the gas is smooth and continuous medium while in the second model it is distributed in a clumpy fashion \citep{2014MNRAS.444L..39L, 2018ApJ...865...70C}. They find that both models fit their data reasonably well. While the smooth continuous model can explain the SED by an injection of relativistic electrons subjected to synchrotron, bremsstrahlung, and ionisation losses, the clumpy model yields negligible losses but a more realistic thermal fractions.

In the dwarf galaxy IC\,10, spectra of several compact H\,{\sc ii}-regions were measured down to 320\,MHz \citep{Basu_2017}. These spectra show evidence of thermal free--free absorption at low frequencies. A region of the edge-on galaxy NGC\,4631 observed by \cite{Stein_2023} is found to have a very flat radio spectrum with a flattening at low-frequencies. Local spectral turnovers are also observed in jellyfish galaxies, where they might be due to ionisation losses in compressed gas \citep{Lal_2022, Ignesti_2022, Roberts_2023}. The only average spiral galaxy where the low frequency effects were studied in detail is M\,51 \citep{Gajovic_2024}. Using high-resolution low-frequency radio maps, they show that the flattening spatially correlates with the neutral H\,{\sc i} gas in the galaxy which points to CR ionisation losses as an important mechanism aside from free--free absorption. 

In summary, there is still no consensus about what causes the low-frequency flattening at radio frequencies below 300 MHz and whether there is a turnover. The most commonly suggested explanation for the flattening is free-free absorption. However, ionisation losses, CR ageing, or synchrotron-self absorption could also be responsible. The physical mechanism most likely depends on the galaxy's environment and is possibly a combination of multiple effects. We therefore choose to explore the low frequency spectrum of the nearby spiral galaxy M\,101 that is know to have multiple giant \hii regions in a diffuse disc. While the previous study of M\,51 surprisingly shows no correlation with ionised gas \citep{Gajovic_2024}, this may be because the H\,{\sc ii} regions are embedded in strong background radiation and thus are hard to study individually. Therefore, we focus on the giant H\,{\sc ii} regions in M\,101 that are very massive and luminous, so they can be studied in the radio continuum \citep{Israel_1975}. 

The galaxy M\,101 (details in Table \ref{tab:M101}) and its \hii regions have already been studied at radio frequencies down to 610 MHz \citep{Israel_1975, Graeve_1990, Berkhuijsen_2016}. It is possible to observe M\,101 at a high spatial resolution because of its large angular diameter of $40\arcmin$ and a distance of only 6.6 Mpc. \cite{Israel_1975} mapped M\,101 at 610 MHz with $1\arcmin$ resolution, and specifically noted the brightest \hii regions and how the structure visible in radio compares to two different H$\alpha$ maps. \cite{Graeve_1990} studied the \hii region spectra and found them to be flat due to thermal emission. However, they did not observe a turnover because their measurements did not extend to frequencies below 610 MHz. In this work, we use LOFAR to observe M\,101 at frequencies of 54 and 144 MHz with a resolution higher than $26\arcsec$ and search for the low frequency flattening in the \hii regions.

\begin{table}
        \caption{Basic properties of M\,101.}
        \label{tab:M101}
        \centering
        \begin{tabular}{@{} lc @{}}
                \toprule
                \toprule
                Name & M\,101 \\
                Alternative names & NGC\,5457 \\
                RA (J2000) & 14$^{\mathrm{h}}$03$^{\mathrm{m}}$12\fs5\\
                DEC (J2000) & +54\degr20\arcmin56\arcsec \\
                Morphology &  	SAB(rs)cd\tablefootmark{a} \\
                Apparent size & $40\arcmin$ \tablefootmark{b} \\
                Distance & $6.644 \pm0.087$\,\text{Mpc}\tablefootmark{c} \\
                Position angle & 36\degr \tablefootmark{d} \\
                Inclination angle & 26\degr \tablefootmark{d} \\
                Scaling  & 1$\arcsec \leftrightarrow$ 32.21 pc \\
                \bottomrule
        \end{tabular}
        \tablebib{
        \tablefoottext{a}{\citet{deVaucouleurs_1991}} \tablefoottext{b}{\citet{Vorontsov_1962}} \tablefoottext{c}{\citet{Hiramatsu_2023}} \tablefoottext{d}{\citet{Hu_2018}} }
\end{table}

This paper is structured as follows: in Sect.~\ref{section_data} we present the data from LOFAR, Apertif and literature. Sect.~\ref{section_analysis} describes the data preparation, including the selection of the \hii regions. In Sect.~\ref{section_results} we present our results that are discussed in Sect.~\ref{section_discussion}. Finally, we conclude in Sect.~\ref{section_conclusions}.

\section{Data}
\label{section_data}

We used four radio maps at frequencies of 54, 144, 1370 and 4850 MHz at resolutions between $11\arcsec$ and $30\arcsec$.
The details of each map are presented in Table \ref{tab:data} and in the following sections.

\begin{table*}
\begin{minipage}{\textwidth}
	\caption{Observation parameters for the radio maps used in the analysis. }
	\label{tab:data}
	\centering
	\newcolumntype{0}{>{\centering\arraybackslash} m{2.3cm} }
	\newcolumntype{1}{>{\centering\arraybackslash} m{1.2cm} }
	\newcolumntype{2}{>{\centering\arraybackslash} m{2.0cm} }
	\newcolumntype{3}{>{\centering\arraybackslash} m{1.5cm} }
	\newcolumntype{4}{>{\centering\arraybackslash} m{1.5cm} }
	\newcolumntype{5}{>{\centering\arraybackslash} m{1.5cm} }
	\newcolumntype{6}{>{\centering\arraybackslash} m{1.5cm} }
    \newcolumntype{7}{>{\centering\arraybackslash} m{1cm} }
	\renewcommand{\arraystretch}{1.5}
	\begin{tabular}{@{} 0 1 2 3 4 5 6 7 @{}}
		\toprule
		\toprule
		Telescope (Survey) & Central frequency [MHz] & Beam size & Beam position angle PA [\degree] & Original $\sigma$ rms noise [$\upmu$Jy\,beam$^{-1}$] & Convolved regridded $\sigma_c$ rms noise [$\upmu$Jy\,beam$^{-1}$] & Flux density scale uncertainty $\epsilon_\nu$ [\%] & Ref. \\
		\midrule
        LOFAR (LoLSS) & 54 & $26\farcs 5\times 14\farcs 7$ & 114.8 & 1700 & 1700 & 10 & 1 \\
        LOFAR (LoTSS) & 144 & $20\farcs0 \times 20\farcs 0$ & / & 130 & 170 & 10 & 2 \\
		Apertif & 1370 & $15\farcs 1\times 11\farcs 0$ & 1.4 & 20 & 30 & 5 & 3 \\
		VLA+Effelsberg & 4850 & $30\farcs 0\times 30\farcs 0$ & / & 59 & 59 & 5 & 4 \\
		\bottomrule
	\end{tabular}
    \tablebib{(1) Observed by \citet{deGasperin_2021}, reprocessed in this paper; (2) \citet{Shimwell_2022}; (3) This work; (4) \cite{Wezgowiec_2022};}  \\
\end{minipage}
\end{table*}

\subsection{LOFAR LBA observations}
\begin{figure}
	\resizebox{\hsize}{!}{\includegraphics{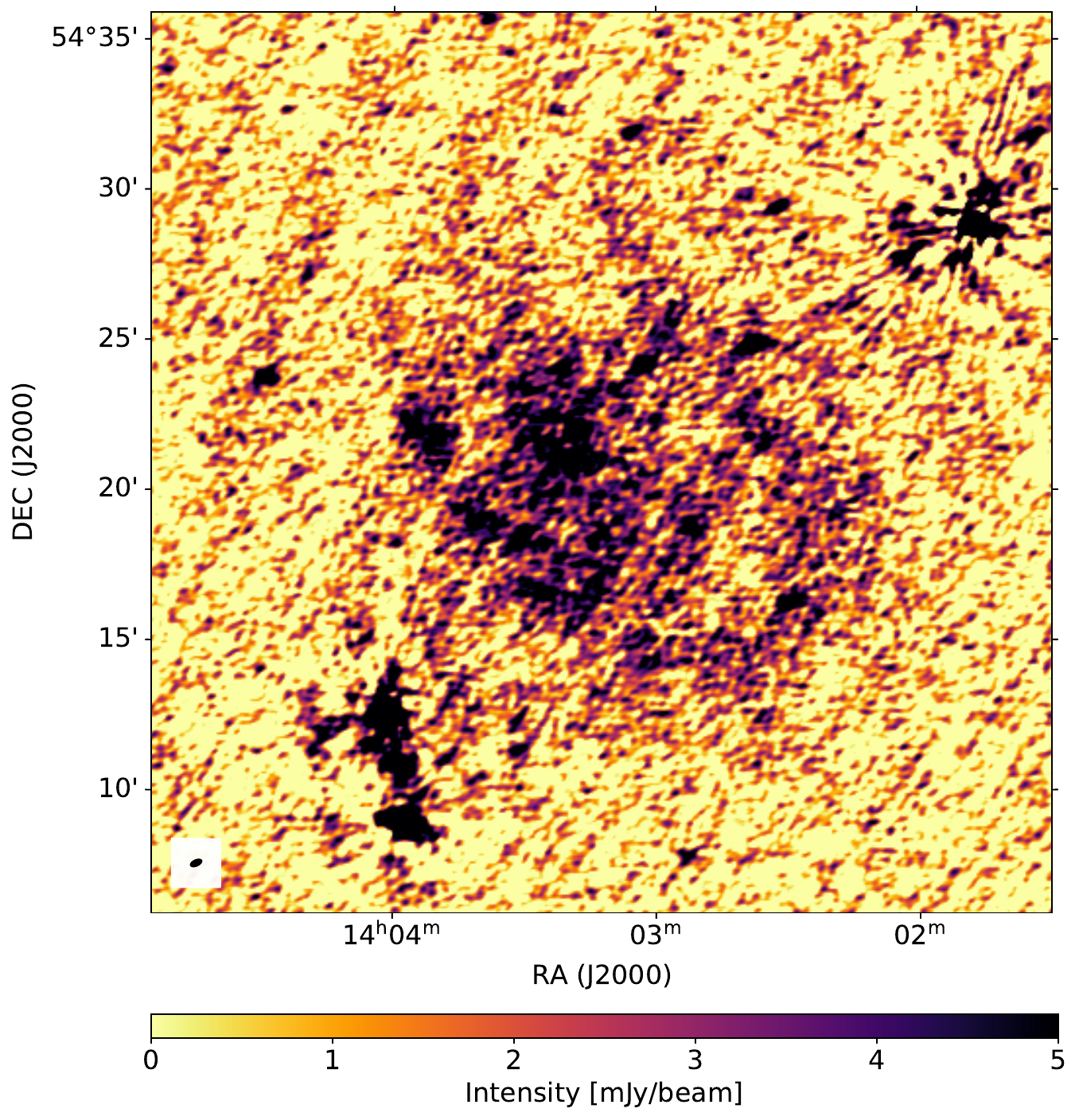}}
	\caption{Radio map of M\,101 at 54\,MHz observed with LOFAR LBA. The beam size is shown with the black ellipse in the bottom left corner.}
	\label{fig:LBA_map}
\end{figure}

M\,101 was observed with the LOFAR Low-Band Antenna (LBA) system at a central frequency of 54\,MHz and in a bandwidth of 24\,MHz as part of the LOFAR LBA Sky Survey \citep[LoLSS;][]{deGasperin_2021, deGasperin_2023}. The galaxy is covered by three survey pointings (P209+55, P210+52, P214+55) with a primary beam response of at least 30\% and each pointing was observed for 8\,h. The observations were calibrated for direction-independent \citep{deGasperin_2021} and direction-dependent effects \citep{deGasperin_2023}. To further process the pointings, for each observation we subtracted all sources except for a circular region with $0.25^\circ$ radius around M\,101 from the visibility data using the source models and calibration solutions found in the direction-dependent calibration carried out as part of LoLSS. We then phase-shifted the measurement sets of the individual observations and pointings to the location of M\,101 and corrected them for the primary beam response in the new phase centre. We also applied the direction-dependent calibration solutions of the nearest direction-dependent calibrator source from LoLSS \citep[for more details about the calibration procedure see e.g.][]{Pasini_2022}. All nine observations were then imaged together to allow for a deep deconvolution of the extended emission associated with M\,101. The LOFAR LBA radio map of M\,101 at 54\,MHz is presented in Fig.~\ref{fig:LBA_map} and the main parameters of the map are in Table \ref{tab:data}.

\subsection{Apertif observations}

The Apertif system is a Phased Array Feed (PAF) receiver operating at a central frequency of 1.4\,GHz with a bandwidth of 300\,MHz mounted on the dishes of the Westerbork Synthesis Radio Telescope (WSRT). The PAF allows simultaneous observations of 40 individual compound beams covering an area of 6.6\,deg$^2$. 

We inspected the Apertif Wide and Medium Deep Extragalctic Survey \citep[AWES/AMES;][]{Adams_2020} observations and the Apertif archive for coverage of M\,101. Fortunately, the galaxy is situated very close to the intrahour-variable source J1402+5347 \citep{Osterloo_2020} that was observed twelve times during the Apertif operational phase. We considered a compound beam response level of 10\,\% as useful for imaging. This resulted in three compound beams (numbers 29, 35 and 36) of each observation being considered for further inspection.

For the data reduction, we used the cross-calibrated visibilities provided by the \texttt{apercal}-pipeline averaged down to 192 channels for every compound beam data set. These data were then directionally dependent self-calibrated and finally imaged in total power using the advanced Apertif reduction pipeline \citep{Kutkin_2023}.

\begin{figure}
	\resizebox{\hsize}{!}{\includegraphics{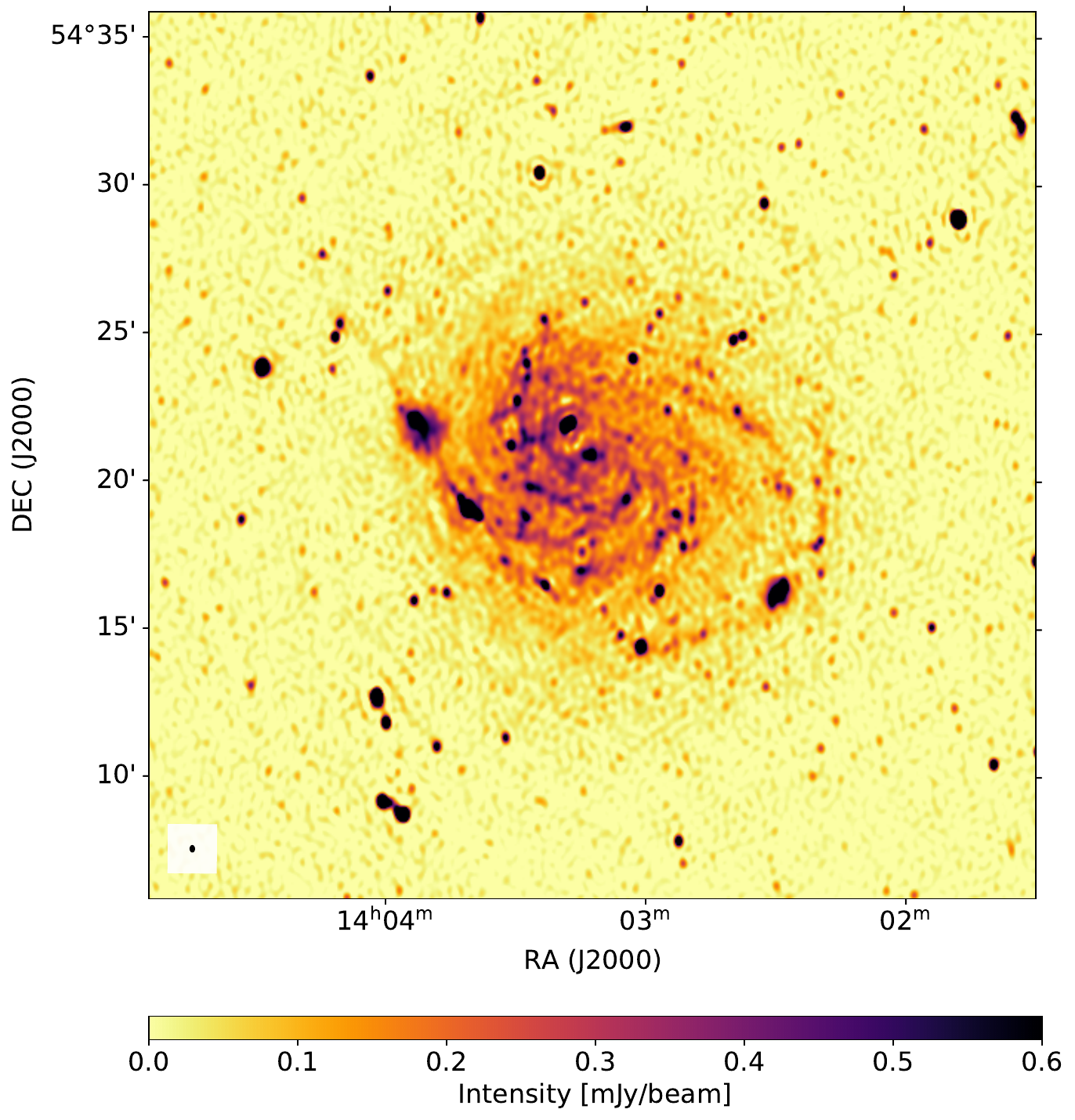}}
	\caption{Radio map of M\,101 at 1370\,MHz observed with Apertif. The beam size is shown with the black ellipse in the bottom left corner.}
	\label{fig:Apertif_map}
\end{figure}

We found that eight out of the twelve observations had sufficient data quality in terms of dynamic range and resolution to be usable for our analysis. The corresponding observation ids are 190807041, 190913045, 191010041, 191102001, 191207034, 200106009, 200128124 and 200302074. The final image was generated by convolving all 24 images (one per beam and observation) to the largest common synthesised beam and correcting by their appropriate primary beam model \citep{Denes_2022}. The final Apertif radio map at 1370\,MHz is presented in Fig.~\ref{fig:Apertif_map} and the main parameters of the map are in Table \ref{tab:data}.

\subsection{Other radio continuum data}

The observation of M\,101 at 144\,MHz was done with the LOFAR high-band antennas (HBA) as a part of the LOFAR Two-metre Sky Survey data release 2 \citep[LoTSS-DR2;][]{Shimwell_2022}. The map of M\,101 used in this work was reprocessed by \citet{Heesen_2022} along with 44 other nearby galaxies. They compared the low $20\arcsec$ and the high $6\arcsec$ resolution maps and made sure that the integrated flux density matches. 
We used the low resolution map because it is more sensitive to diffuse emission and the maps at other frequencies have similar beam sizes, so this map did not have to be convolved to a significantly larger beam. We also consulted the high resolution map to identify background sources. The flux density scale error of LoTSS-DR2 is below 10\% \citep{Shimwell_2022}, which is the value we adopted. The main parameters of the LOFAR HBA radio map are presented in Table \ref{tab:data}.

The radio map of M\,101 at 4850 MHz is a combination of VLA and Effelsberg maps presented by \cite{Wezgowiec_2022}. Because of the large angular size of M\,101, the VLA observations include multiple pointings separated in two observation projects (19B-027 and 15B-292). The results of Effelsberg observations were previously published by \cite{Berkhuijsen_2016}. The combination of the two data sets restores the large-scale diffuse emission that cannot be recovered if the short baselines are not sufficiently covered by the interferometer. 
More details on the VLA calibration and the combination of data sets are presented in \cite{Wezgowiec_2022}. In this work, we used only the total intensity map and assumed a flux density scale uncertainty of 5\%, which is standard for the VLA \citep{Konar_2013}. The main parameters of the combined VLA and Effelsberg radio map are presented in Table \ref{tab:data}.

\section{Data preparation}
\label{section_analysis}

The radio maps were additionally processed in the following way. All data, except for the 4850\,MHz, were convolved to the largest common beam of $26.52\arcsec\times20\arcsec$ with the position angle PA=115\degr. For the convolution, we used the routine \texttt{CONVOL} of the software package \texttt{MIRIAD} \citep[][]{Sault_1995}. The 4850\,MHz data were kept with the original 30$\arcsec$ beam and for flux density measurements we made sure to take into account the different beam size for that map. We did not convolve to the largest common beam of all data (which would be 30\arcsec) because that would make the large rms noise in the LBA map even worse. The rms error dominates over the flux scale error in small regions that we study, so it was important to keep it as small as possible.

All maps were then aligned to a common world coordinate system, re-gridded to a pixel size of $2\arcsec$, as well as transformed into the reference pixel and image size. We used the world coordinate system of the H$\alpha$ map by \cite{Hoopes_2001} as the reference. This was achieved using the \texttt{REGRID} routine of \texttt{MIRIAD}. 

The rms noise $\sigma$ of each individual image was determined by calculating the standard deviation over an emission-free area. The total noise was calculated as a combination of the rms noise and the flux density scale uncertainty (both listed in Table \ref{tab:data}). 

\subsection{Data validation}
\label{section_validation}

We first checked if the integrated flux density of M101 in our maps matches the literature measurements. We prepared a mask to calculate the integrated flux density in our maps, which includes all the emission from the galaxy, but excludes nearby bright radio sources (e.g. NVSS J140147+542852 and NVSS J140357+540852). We specifically made sure we include the \hii region NGC 5471, which is located outside the galactic disc to the east. We did not remove the background sources in cases where their emission overlapped with the disc of the galaxy (e.g. NVSS J140318+542159). The mask is shown in Appendix \ref{a:mask_int}.

\begin{figure}
	\resizebox{\hsize}{!}{\includegraphics{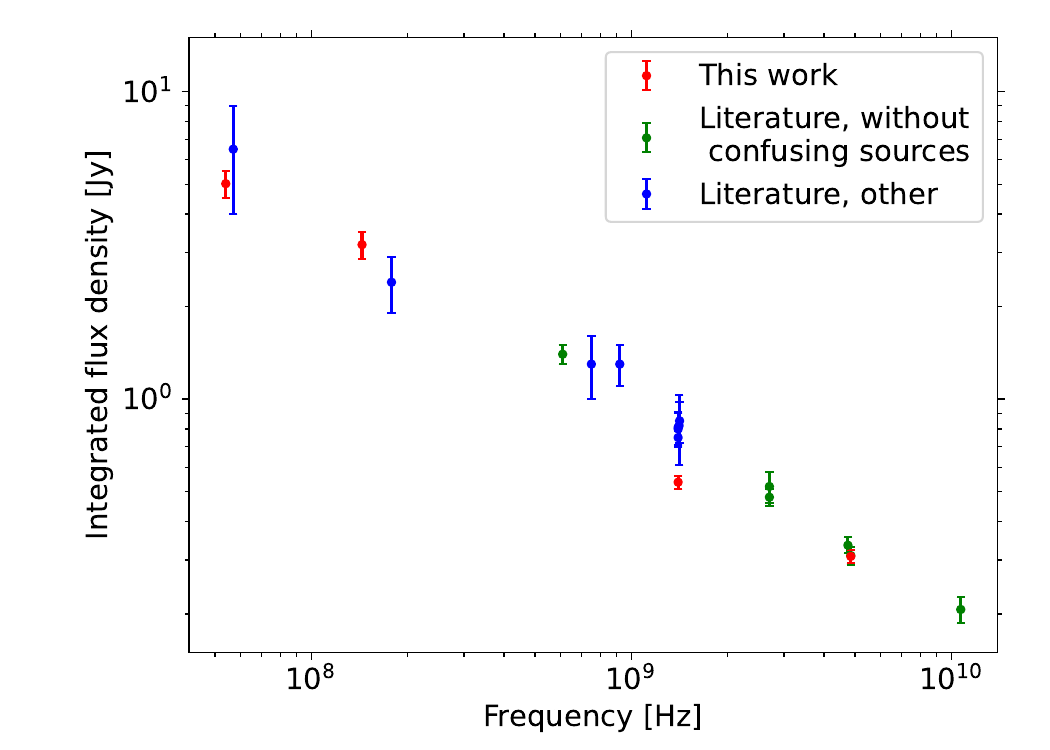}}
	\caption{Integrated spectrum of M\,101. The flux density measurements from the maps used in this work (red) compared to literature flux density measurement, separated into those where bright background sources surrounding the galaxy are excluded (green) and are not excluded (blue). 
 }
	\label{fig:integrated}
\end{figure}

We compile the integrated flux density of M\,101 from the literature and from the maps used in this paper in Appendix~\ref{a:integrated}. We also note the type of telescope used for the observation, the beam size and most importantly whether the bright background sources around the galaxy are included in the flux density measurements. We show these measurements in Fig.~\ref{fig:integrated}. The data points roughly follow a power-law. We do not see a significant flattening of the integrated spectrum at low frequencies. Comparing our data with the literature, we notice that there might be missing large-scale flux density in the 1370\,MHz Apertif map. The inclusion of background sources cannot account for this difference. For this reason, the Apertif map was not used for studying the galaxy as a whole, only the compact \hii regions that are less likely to be affected by missing flux density issue.

\begin{figure}
	\resizebox{\hsize}{!}{\includegraphics{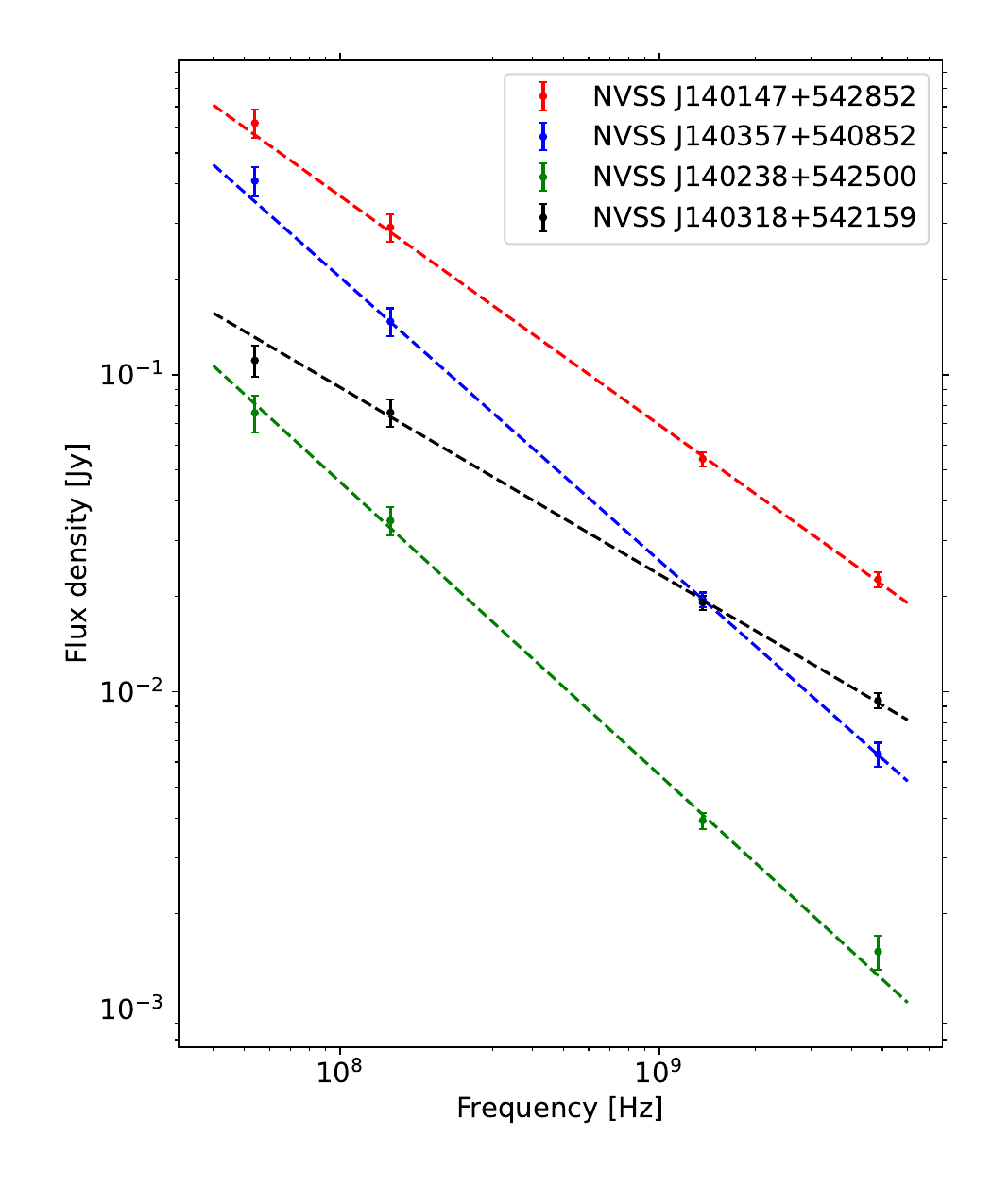}}
	\caption{ Spectra of four selected background sources. The spectra follows a power law as expected confirming that there are no flux density problems in our images at small scales.}  
	\label{fig:point}
\end{figure}

We additionally checked if our maps are reliable to study spectra in smaller regions. In Fig.~\ref{fig:point}, we plot the spectra of four background radio sources surrounding M\,101. The spectra are well-fit by a power-law between 54 MHz and 5GHz, and the offset of the data points is minimal. Because the small scale flux is reliable, we can continue with analysing the spectra of \hii regions in M\,101. We used the Apertif map for studying the spectra of smaller regions that are not affected by the missing large-scale flux density.

\subsection{H {\sc ii} region selection}
\label{section_SI_SC}

More than a thousand \hii regions were identified in M\,101 \citep{Hodge_1990}, but we want to focus on the prominent giant \hii regions. There is no consensus in literature on what constitutes a giant \hii region, but they generally have a large physical size and they are very bright at both optical and radio frequencies. Therefore, we combine the most comprehensive lists of giant \hii regions in M\,101 \citep{Skillman_1988,Hippelein_1986,Dodorico_1983} with coordinate positions \citep{Gordon_2008} in Table \ref{tab:HII_regions}. 

\begin{table}
        \caption{Coordinates of \hii regions considered for analysis. Regions 0-5 were selected.}
        \label{tab:HII_regions}
        \newcolumntype{0}{>{\centering\arraybackslash} m{0.8cm} }
        \newcolumntype{1}{>{\centering\arraybackslash} m{1.9cm} }
        \newcolumntype{2}{>{\centering\arraybackslash} m{1.5cm} }
        \newcolumntype{3}{>{\centering\arraybackslash} m{1.5cm} }
        \newcolumntype{4}{>{\centering\arraybackslash} m{1.5cm} }
        \renewcommand{\arraystretch}{1.5}
        \begin{tabular}{@{} 0 1 2 3 4 @{}}
                \toprule
                \toprule
                No. & Name & RA(J2000) & DEC(J2000) & Ref. \\
                \midrule
        0 & Centre & 14:03:12.48 & +54:20:55.4 	 &  1 \\
        1 & NGC 5447 & 14:02:28.18 & +54:16:26.3  &  1,2,3,4 \\
        2 & NGC 5455 & 14:03:01.13 & +54:14:28.7  &  1,2,3,4 \\
        3 & NGC 5461 & 14:03:41.36 & +54:19:04.9  &  1,2,3,4  \\
        4 & NGC 5462 & 14:03:53.19 & +54:22:06.3  &  1,2,4  \\
        5 & NGC 5471 & 14:04:29.35 & +54:23:46.4  &  1,2,3,4 \\
        6 & NGC 5449 & 14:02:26.0  & +54:19:48.0  &  2,5 \\
        7 & Hodge 1013\tablefootmark{a, b} & 14:03:31.0  & +54:21:14.5  &  1,2,4,6 \\
        8 & Searle 5 & 14:02:55.05 & +54:22:26.6  &  1,2 \\
        9 & Searle 12 & 14:04:11.11 & +54:25:17.8  &  1,2,4 \\
        10 & Hodge 67\tablefootmark{a} & 14:02:19.92 & +54:19:56.4  &  1 \\
        11 & Hodge 70/71\tablefootmark{a} & 14:02:20.50 & +54:17:46.0  &  1 \\
        12 & Hodge 681\tablefootmark{a} & 14:03:13.64 & +54:35:43.0  &  1 \\        
                \bottomrule
        \end{tabular}
        \tablebib{(1) \cite{Gordon_2008}; (2) \citet{Hippelein_1986}; (3) \cite{Skillman_1988}; (4) \cite{Dodorico_1983}; (5) \cite{Corwin_1995}; (6) \cite{Hodge_1990}; \\}
        \tablefoottext{a}{Identification numbers according to \cite{Hodge_1990}. }
        \tablefoottext{b}{Also called Hodge 40, according to \cite{Hodge_1969}. }
        
\end{table}

\begin{figure}
	\resizebox{\hsize}{!}{\includegraphics{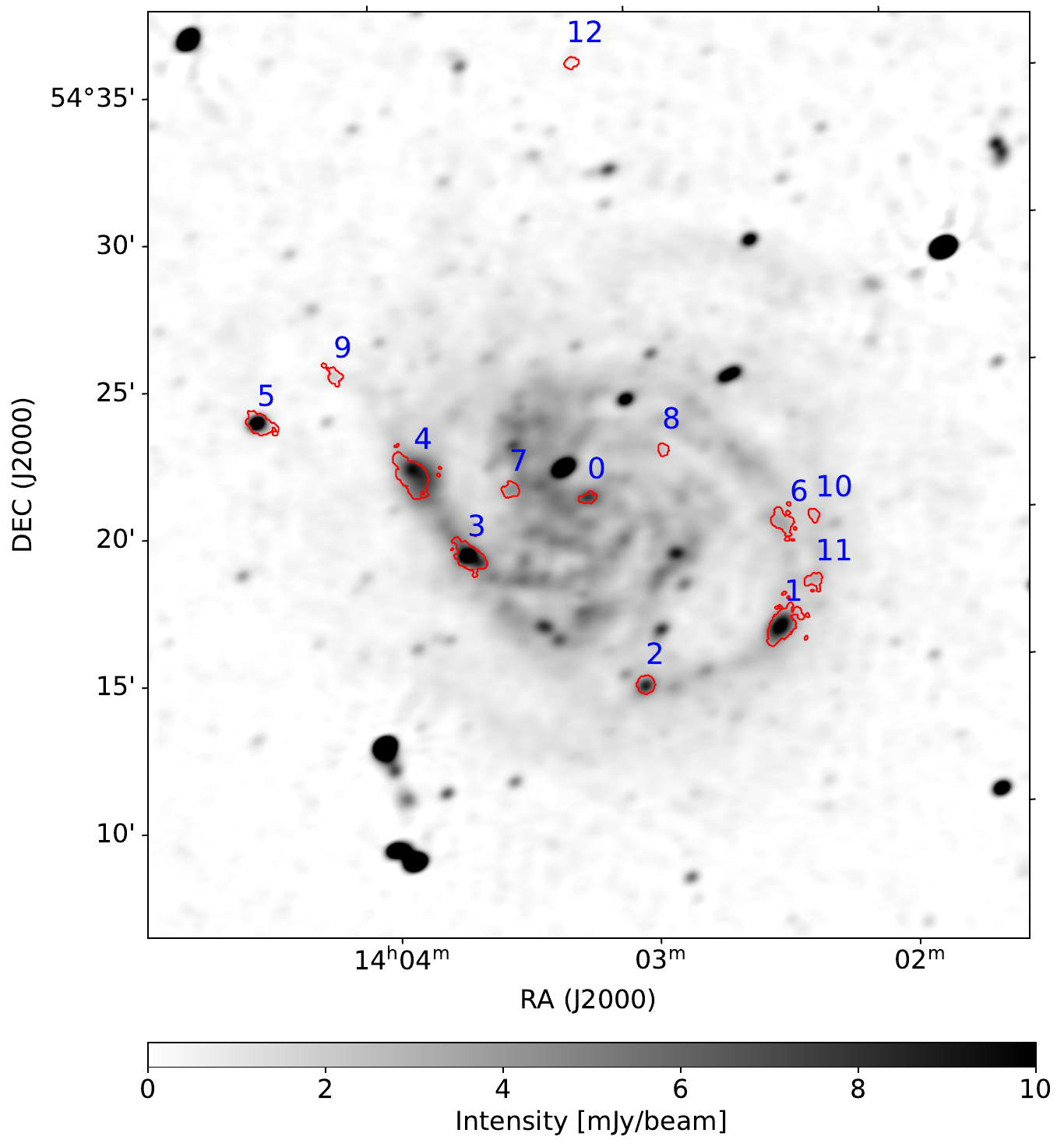}}
	\caption{ \hii region boundaries determined from H$\alpha$ \citep{Hoopes_2001} plotted on top of 144\,MHz radio map. The region boundaries are red and the blue numbers correspond to their identification numbers in Table \ref{tab:HII_regions}. For the analysis we select or only regions 0-5, which are bright at 144\,MHz. Note that these boundaries were only used in the \hii region selection.}
	\label{fig:region_selection}
\end{figure}

For this paper, we selected the \hii regions that are prominent at low radio frequencies, specifically in the 144\,MHz map. We chose the 144\,MHz map because the frequency is high enough to not be affected by low frequency flattening. Additionally, the 144\,MHz map has a good signal-to-noise ratio. In this selection, we used the H$\alpha$ map by \cite{Hoopes_2001} to define the \hii region boundaries. First, we manually plotted circles that include all the H$\alpha$ emission from the \hii region and mostly exclude the surrounding emission. Second, we refined the \hii region boundary by selecting everything above 10$\sigma$ in the H$\alpha$ image. 
In Fig.~\ref{fig:region_selection}, the \hii region boundaries defined from H$\alpha$ are plotted on top of the 144 MHz map. Regions 0--5 show a clear excess of emission compared to the disc surrounding them. For regions 0--5, surface brightness is higher than 7 mJy/beam while for the other regions it is lower than 5 mJy/beam. We chose these six regions (the galactic centre, NGC 5447, NGC 5455, NGC 5461, NGC 5462, and NGC 5471) for the spectral analysis in this work. Coincidentally, those are the same regions selected for analysis by \cite{Israel_1975}, likely because of their brightness at radio frequencies.

\section{Results}
\label{section_results}

\subsection{Spectral index maps between 54 and 144 MHz}

We calculated the radio spectral index of M\,101 between 54 and 144\,MHz. Both radio maps were convolved to the same beam ($26.52\arcsec\times20\arcsec$ beam with PA=115\degr) and re-gridded to the same coordinate grid. The spectral indices were calculated in a standard way by taking the logarithmic ratio of the flux densities in two maps over the logarithmic frequency ratio. The errors in the spectral index maps were calculated by propagating the errors from the individual maps, which are a combination of the flux density scale uncertainty and the rms noise measured away from the source (both listed in Table \ref{tab:data}). The regions below 2$\sigma$ (in ether map) were excluded from the analysis. We chose a 2$\sigma$ exclusion limit instead of the usually used 3$\sigma$ because otherwise the high noise in the LBA map would significantly limit the area in which the spectral index is calculated. We also note that LOFAR has good coverage of short baselines which correspond to large angular scales on the sky. We can therefore trust that the LOFAR maps are not missing large scale emission from M\,101 and the spectral indices presented here are reliable.

The index of M\,101 between 54 and 144 MHz is shown in Fig.~\ref{fig:spix} and the corresponding error map can be found in Appendix~\ref{a:error_map}. Significant areas of flat or inverted spectra index are present, especially at the positions of the \hii regions. 

\begin{figure}
	\resizebox{\hsize}{!}{\includegraphics{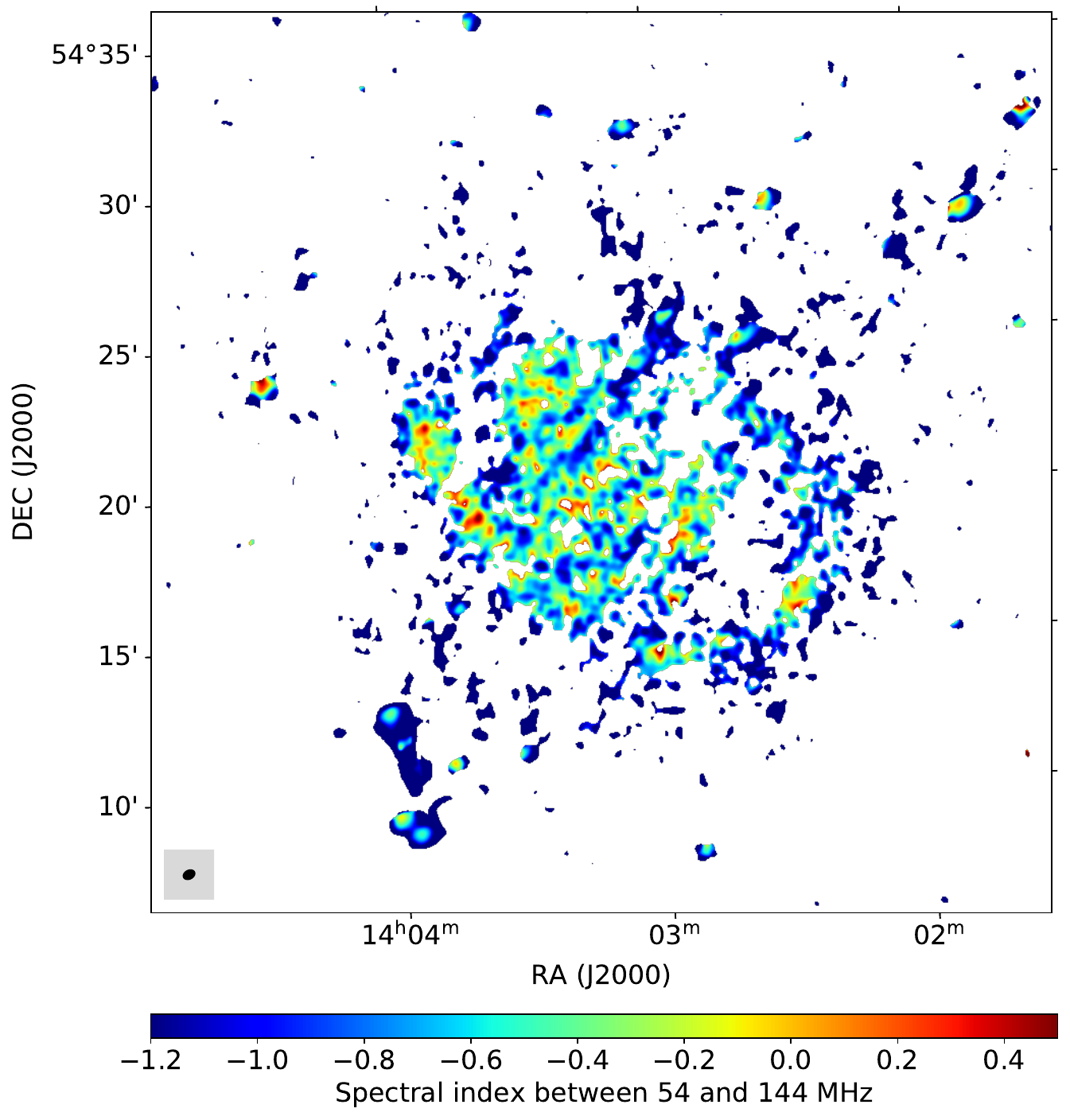}}
	\caption{ Spectral index map of M\,101 between 54 and 144\,MHz. White areas below 2$\sigma$ are excluded. The beam size is shown with the black ellipse in the bottom left corner. }
	\label{fig:spix}
\end{figure}

\begin{figure*}[!h]
    \centering
    \resizebox{\hsize}{!}{\includegraphics{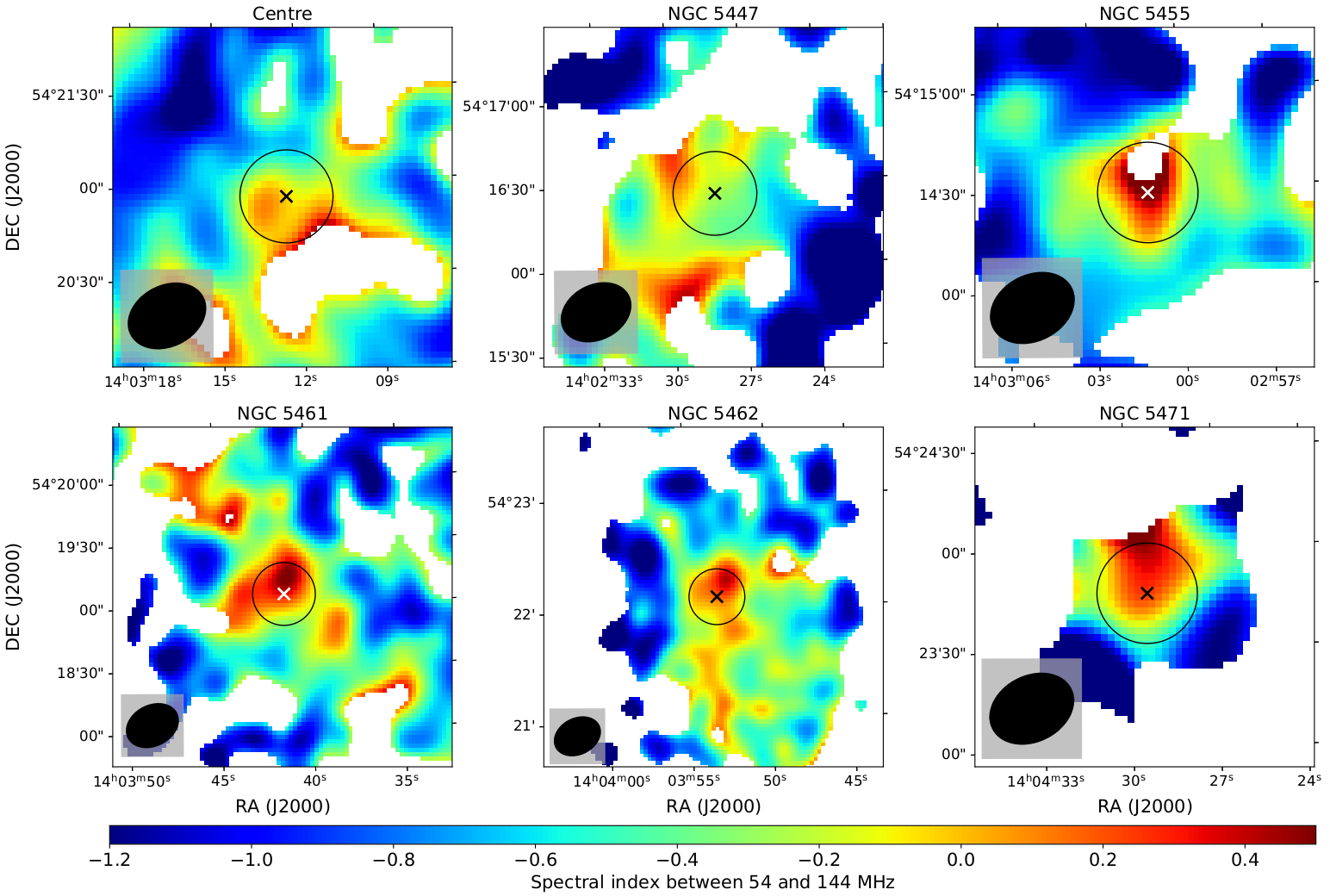}}
    \caption{ Zoomed-in spectral index maps of \hii regions in M\,101 between 54 and 144\,MHz. The regions where the spectrum is measured (in Sect. \ref{section_SEDs}) are indicated with black circles and the centres of the regions (Table \ref{tab:HII_regions}) are marked with an x. White areas below 2$\sigma$ are excluded. The beam size is shown with the black ellipse in the bottom left corner of each panel. 
    }
    \label{fig:zoom_ins}
\end{figure*}

In Fig.~\ref{fig:zoom_ins} we present cutouts of the spectral index maps focused on the six \hii regions that we selected for this study. For the majority of the regions we see a small area of positive spectral index right at the centre of several \hii regions. There, we expect the highest free--free absorption because of high density of ionised gas. If we want to observe the turnover in the spectrum, we should focus on these very central areas.

\subsection{SEDs of H {\sc ii} regions and fits}
\label{section_SEDs}

\begin{figure*}[!h]
    \centering
    \resizebox{\hsize}{!}{\includegraphics{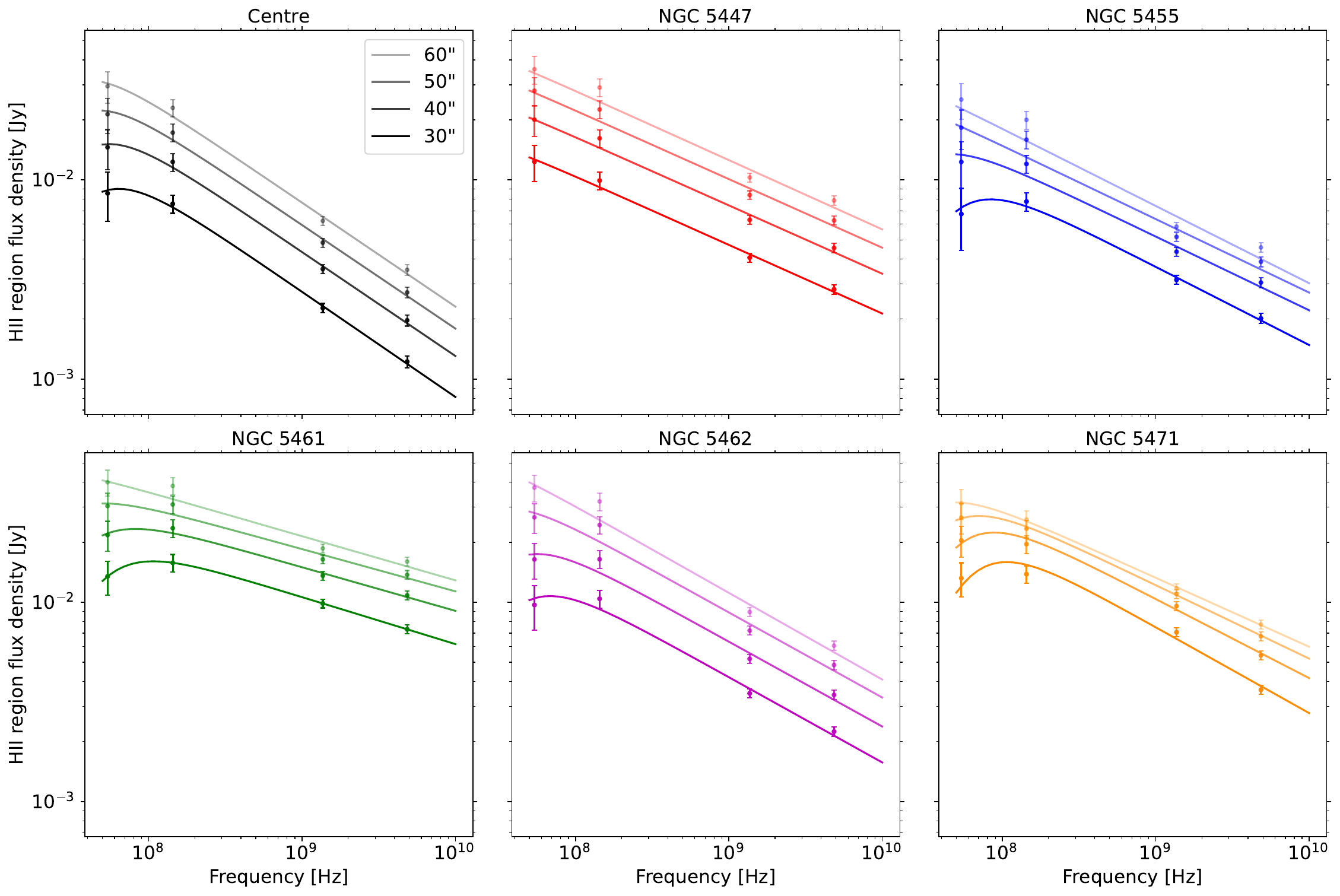}}
    \caption{ Spectra of the selected \hii regions in M\,101. The solid lines of different opacity correspond to fits for the free-free absorption model (combined Eq. \ref{equation_thermabs} and \ref{equation_opacity}) in different areas within which the flux is measured, with diameters from 30 to 60$\arcsec$ with a 10$\arcsec$ step. The fit parameters for the 30$\arcsec$ area are in Table \ref{tab:fit_res}. 
    }
    \label{fig:SEDs}
\end{figure*}

In the spectral index cutouts (Fig.~\ref{fig:zoom_ins}) we see that there are areas of positive spectral index at the very centre of \hii regions (except NGC\,5447). For this reason we measure the spectra in circular areas with increasing size starting from the largest beam size (30\arcsec) up to twice the beam size (60\arcsec) with a 10\arcsec step. These areas are centred on the position of the \hii region according to \cite{Gordon_2008}, which are listed in Table \ref{tab:HII_regions}. Alternative centres could be defined directly from H$\alpha$ data, but the position would not change more than the size of our beam and an alternative definition would not affect our results significantly. Here, we include measurements at all four frequencies (54, 144, 1370, and 4850 MHz) because the flux densities in small regions are accurate in all the maps, which we confirm by checking the spectra of background sources in Sect \ref{section_validation}.

The spectra are shown in Fig.~\ref{fig:SEDs} with different lines indicating different areas within which the fluxes are measured. We can clearly see a deviation from a power law with decreasing area, or increasing resolution. We initially model the spectra using a synchrotron power law with an internal free--free absorbing medium. We make an assumption that in the \hii region the absorbing thermal plasma coexists with the synchrotron-emitting CR electrons. Thus, the model assumes a thoroughly mixed medium of thermal gas and cosmic rays. This "internal" model includes emission and free-free absorption internal to a continuous synchrotron-emitting region. As in \cite{2017ApJ...838...68K}, the model only considers the non-thermal synchrotron emission and the free--free absorption of that emission. At such low frequencies, the free--free emission is insignificant compared to the synchrotron emission so the free--free self-absorption is neglected and free--free absorption only affects synchrotron emission. Modelling the radio spectrum with a synchrotron power law with an internal free--free absorbing screen, the flux density $S_\nu$ depends on frequency as \citep{Tingay_2003}: 

\begin{equation}
        S_\nu = S_0\ \bigg(\frac{\nu}{\nu_0} \bigg)^{\eta}\ \bigg( \frac{1- e^{-\uptau_{\rm ff}}}{\uptau_{\rm ff}} \bigg)\ ,
        \label{equation_thermabs}
\end{equation}
where $S_0$ is the flux density normalisation, $\nu$ is the frequency in GHz, $\nu_0$=1\,GHz is the reference frequency, $\eta$ is the absorption--free spectral index and $\uptau_{\rm ff}$ is the free--free emission optical depth. Here, the last term in the bracket on the right-hand side of the equation denotes the escape probability of the photons; this expression can be obtained by solving the radiation transport equation using the large-scale velocity approximation under the assumption of spherical symmetry \citep{deJong_1975}. In the expression for the free--free Gaunt factor at radio wavelengths from \cite{RevModPhys.34.507}, the optical depth is given as

\begin{equation}
 \uptau_{\rm ff} = {8.2 \times 10^{-2}\, \nu^{-2.1}\, {T_{\rm e}^{-1.35}}\, {\rm EM}},
        \label{equation_opacity}
\end{equation}
with the emission measure (EM) in units of cm$^{-6}$\,pc and the electron temperature $T_{\rm e}$ in K. The EM is defined as 
\begin{equation}
        \left(\frac{\mathrm{EM}}{\rm pc\,cm^{-6}}\right) = \int_{0}^{s_0} \left( \frac{n_e}{\rm cm^{-3}} \right)^2\ \left( \frac{{\rm d}s}{\rm pc} \right),
\end{equation}
where $n_e$ is the thermal free electron density, and d$s$ is the infinitesimal depth element along the line of sight between the source at $s_0$ and an observer at 0. We combine equations \eqref{equation_opacity} and \eqref{equation_thermabs} to make a single free--free absorption fit model with three free parameters: $S_0$, $\eta$ and EM. The electron temperature is assumed to be $T_e=10^4$ K. The value of electron temperature does not affect the shape of the spectrum, only the resulting value of EM.

Alternatively, the free--free absorption could occur in the intervening region (essentially a screen), which is also likely to contain thermal gas, in particular since the \hii regions are small compared to the thickness of the disk. Our data cannot distinguish between this external model and the internal model presented in Eq. \ref{equation_thermabs}, both of them fit the data and they look remarkably similar. We use the internal model as our fiducial model. 
In an external model, the spectrum no longer has the optical depth in the denominator, yielding \citep{Wills_1997}

\begin{equation}
        S_\nu = S_0\ \bigg(\frac{\nu}{\nu_0} \bigg)^{\eta}\ e^{-\uptau_{\rm ff}} .
        \label{equation_screen}
\end{equation}

The external model leads to lower valued of EM because all the absorption occurs in front of the source thus requiring less ionised gas. The fit results for both models are in Table \ref{tab:fit_res}.

\begin{table*}[t]
        \caption{ Fitted parameters for the spectra of the giant \hii regions in M\,101. The spectra are measured in areas with a 30$\arcsec$ diameter. }
        \label{tab:fit_res}
        \centering
        \newcolumntype{0}{>{\centering\arraybackslash} m{3cm} }
        \newcolumntype{1}{>{\centering\arraybackslash} m{3cm} }
        \newcolumntype{2}{>{\centering\arraybackslash} m{3cm} }
        \newcolumntype{3}{>{\centering\arraybackslash} m{3cm} }
        \newcolumntype{4}{>{\centering\arraybackslash} m{3cm} }
        \renewcommand{\arraystretch}{1.5}
        \begin{tabular}{@{} 0 1 2 3 4 @{}}
                \toprule
                \toprule
                &  \multicolumn{2}{c}{Internal model} &  \multicolumn{2}{c}{ External model} \\
                Region & $\eta$ & EM [pc cm$^{-6}$] & $\eta$ & EM [pc cm$^{-6}$] \\
                \midrule
Centre & $-0.528\pm 0.044$ & $5200\pm 5200$ & $-0.528\pm 0.043$ & $2500\pm 2200$ \\
NGC 5447 & $-0.344\pm 0.035$ & $200\pm 3300$ & $-0.343\pm 0.039$ & $0\pm 1800$ \\
NGC 5455 & $-0.393\pm 0.041$ & $6700\pm 6100$ & $-0.392\pm 0.040$ & $3200\pm 2600$ \\
NGC 5461 & $-0.236\pm 0.036$ & $6600\pm 4100$ & $-0.235\pm 0.035$ & $3000\pm 1700$ \\
NGC 5462 & $-0.428\pm 0.041$ & $4800\pm 4600$ & $-0.429\pm 0.040$ & $2300\pm 2000$ \\
NGC 5471 & $-0.431\pm 0.036$ & $12300\pm 5300$ & $-0.424\pm 0.034$ & $4900\pm 1800$ \\
                \bottomrule
        \end{tabular}
\end{table*}

In most of the analysed \hii regions the flux density at 54 MHz is significantly lower than it would be if the power law behaviour extended to lower frequencies. The exception is the region NGC\,5447 where the spectrum is entirely a power law with no absorption at low frequencies. We exclude this region from further collective analysis and analyse it separately in Sect.~\ref{section_reg1}.

We observe a flattening in the spectrum of \hii regions at frequencies below 100 MHz. 
Assuming $T_e=10^4$ K and the "internal" free-free absorption model we get values of the EM between 4800 and 12000 cm$^{-6}$\,pc. Fitting a different temperature does not change the fits because the parameter $T_e^{-1.35} \cdot \mathrm{EM}$ enters the fit (Eq. \ref{equation_opacity}). Hence changing $T_e$ leads to a change in the inferred value of EM. \cite{Kennicutt_2003} measured $T_e$ for each of the \hii regions and the values for $T_e$ vary between 9000 K and 12000 K for the low-ionisation zones. This would lead to a change in EM of 25\% at most (which is less than other uncertainties). 
The initial values of EM are of the same order of magnitude but higher than the values determined for the spiral arms of M51 assuming that the flattening in the spectrum is caused by free-free absorption \citep{Gajovic_2024}. In Sect.~\ref{section_EM_halpha} we compare the values of EM obtained by fitting the low frequency radio absorption to a different tracer of EM.

The spectral indices of the \hii regions are very close for both models. Their values are between -0.52 and -0.22 which is fairly flat. Thermal emission has a spectral index -0.1 and non-thermal radio emission has a spectral index higher than -0.7. We observe a combination of thermal and non-thermal emission coming from the centres of \hii regions.

We note that synchrotron self-absorption is not a viable explanation for the spectral behaviour as it requires brightness temperatures well in excess of the ones we measure (see \cite{Gajovic_2024} for an exact calculation). Moreover, ionisation losses of the cosmic-ray electrons do not lead to spectral down-turns but only to a flattening of spectra. Also, since HII regions are already ionised, ionisation losses should not play a major role.

\section{Discussion}
\label{section_discussion}

\subsection{Comparison to other EM tracers}
\label{section_EM_halpha}

One can independently estimate the EM from the H$\alpha$ flux density using the relation \citep{Dettmar_1992, Voigtlander_2013}:

\begin{equation}
\mathrm{EM}= 5 \cdot 10^{17} F_{H\upalpha}/\Omega
\end{equation}
where $F_{H\upalpha}$ is the H$\alpha$ flux density in erg cm$^{-2}$ s$^{-1}$, and $\Omega$ is the collecting area in arcsec$^2$ and EM is in cm$^{-6}$\,pc. 

\begin{figure}
	\resizebox{\hsize}{!}{\includegraphics{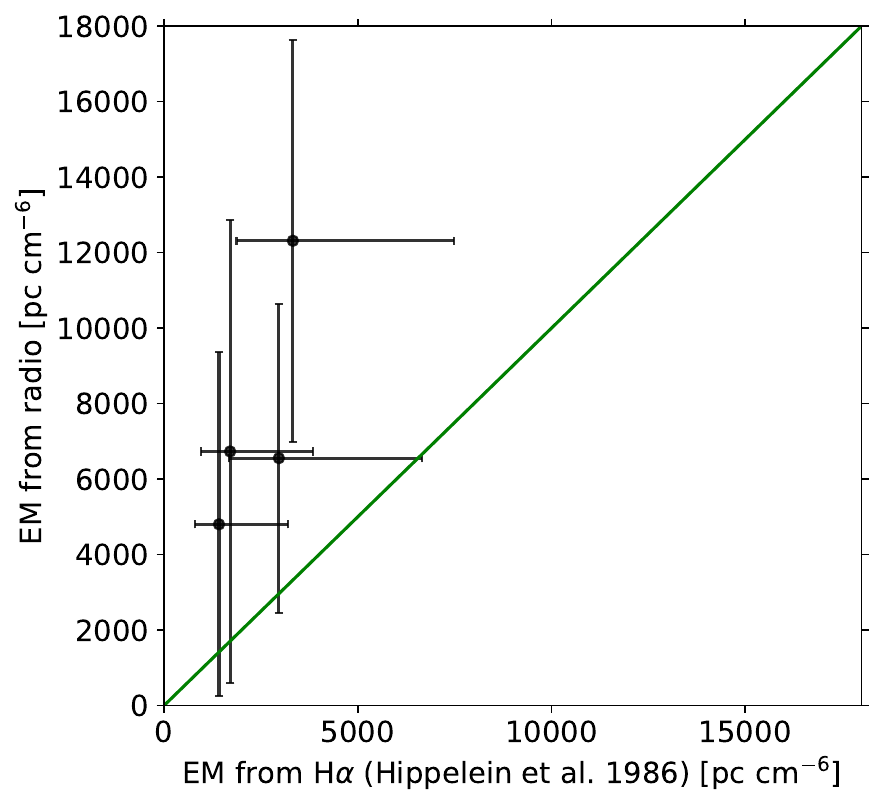}}
	\caption{Comparison of EM values from the internal free-free absorption model fit to the radio data and EM values from H$\alpha$. The 1:1 line is plotted in green. }
	\label{fig:EM_comparison}
\end{figure}

The H$\alpha$ fluxes for our \hii regions (except for the galactic centre) are presented in \citet{Hippelein_1986} who adapted them from \citet{Israel_1980}. The exact collecting areas for each region are not available, but their diameters range between 20\arcsec and 40\arcsec. We assumed a diameter of 30\arcsec for all our regions, because this diameter matches the regions in which we measured the radio flux and the range of possible diameters for the H$\alpha$ measurements. The EM estimate using the H$\alpha$ flux density as proxy is compared to the EM obtained by fitting the internal free-free absorption model to the radio data in Fig.~\ref{fig:EM_comparison}. We exclude the galactic centre because of missing H$\alpha$ data and NGC\,5447 since we do not observe a turnover in that region.

The EM from radio and from H$\alpha$ is the same order of magnitude, but the radio fits are consistently about two times higher then the H$\alpha$ estimates. Here we plot the the EM from the internal free-free absorption model (Eq.~\ref{equation_thermabs}), however in Table~\ref{tab:fit_res} we show that the external model (Eq.~\ref{equation_screen}) results in consistently lower values of EM. The values from the external model are about a factor of two lower, which is more in line with the H$\alpha$. This is an indication that the external model could be more appropriate for the spectra of \hii regions. However, due to the high uncertainty on the radio fit results and the unknown area for H$\alpha$ measurements we cannot firmly claim which model is better.

\subsection{Turnover localisation}

\begin{figure}
	\resizebox{\hsize}{!}{\includegraphics{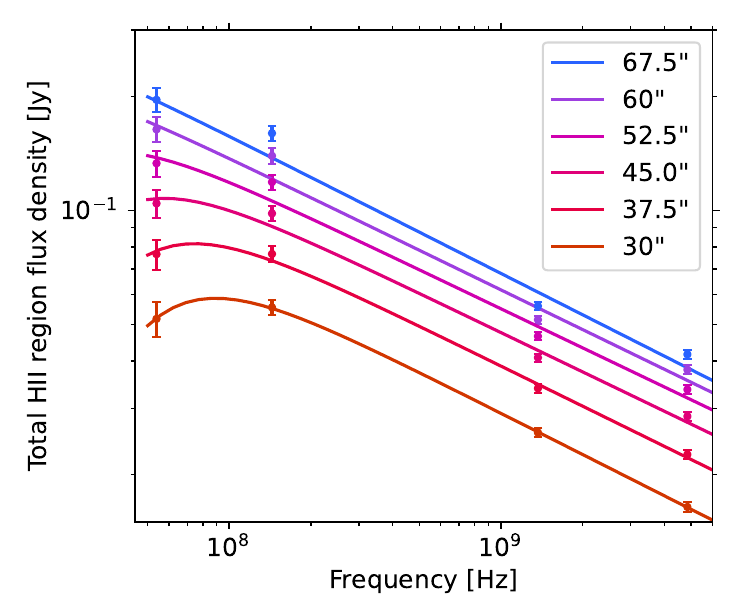}}
	\caption{ Total flux density of \hii regions measured in increasing areas around the centre of the region. NGC\,5447 is excluded.}
	\label{fig:localisation}
\end{figure}

The spectral index maps (Fig.~\ref{fig:zoom_ins}) show inverted spectra right in the centres of the \hii regions. In Fig.~\ref{fig:SEDs}, it is clear that the turnover can only be observed if we focus on a small area around the centre of the \hii region. We now investigate how the size of the area affects our ability to observe a flattening in the spectrum. We added up the fluxes of the five selected \hii regions (excluding NGC\,5447) to create the total \hii region spectrum. To get a mode precise view, we varied the size of the area in which the flux density is measured between 30\arcsec and 67.5\arcsec in steps of 7.5\arcsec. In Fig.~\ref{fig:localisation}, we show how this affects the total \hii region spectrum. We also fitted the free-free absorption model to those spectra. The fluxes of the \hii regions are comparable so the total spectrum is not dominated by the brightest region.

From Fig.~\ref{fig:localisation}, we see that the low frequency flattening in the radio spectrum can not be detected if the area in which the flux density of the \hii region is measured has a diameter larger than 45\arcsec. This corresponds to 1.5\,kpc at the distance of M\,101 (see Table \ref{tab:M101}). This means that the flattening is only observable in the very centre of the \hii region. With the increase of the area we include more of the surrounding part of the \hii region and the flattening gets "diluted". 

It was already shown by \cite{Israel_1975} that more than 50\% (in some regions even up to 90\%) of the radio flux density of the \hii regions in M\,101 is concentrated in the core. It is therefore not surprising that the free-free absorption is also most prominent in the cores, but it is interesting that the area where we detect the flattening has a diameter smaller than 1\arcmin. 

This has important implications for future studies of \hii regions at very low frequencies. To detect the low frequency flattening, the resolution of the radio maps has to be high enough that the centres of the \hii regions with the flattening can be resolved from the surroundings. Our results on M\,101 show that the beam size should be small enough to correspond to 1.5 kpc or less in the observed galaxy. M\,101 has exceptionally bright and dense \hii regions, so we expect that an even better resolution might be needed in other galaxies. The international baselines of the LOFAR telescope allow to reach a resolution around 1\arcsec\ at 54\,MHz \citep{Groeneveld_2022}. If that is combined with SKA observations of similar resolution at higher frequency, in the future it could be possible to observe flattening in extra-galactic \hii regions at distances up to 300\,Mpc.

\subsection{NGC 5447 at low radio frequencies}
\label{section_reg1}

\begin{figure}
	\resizebox{\hsize}{!}{\includegraphics{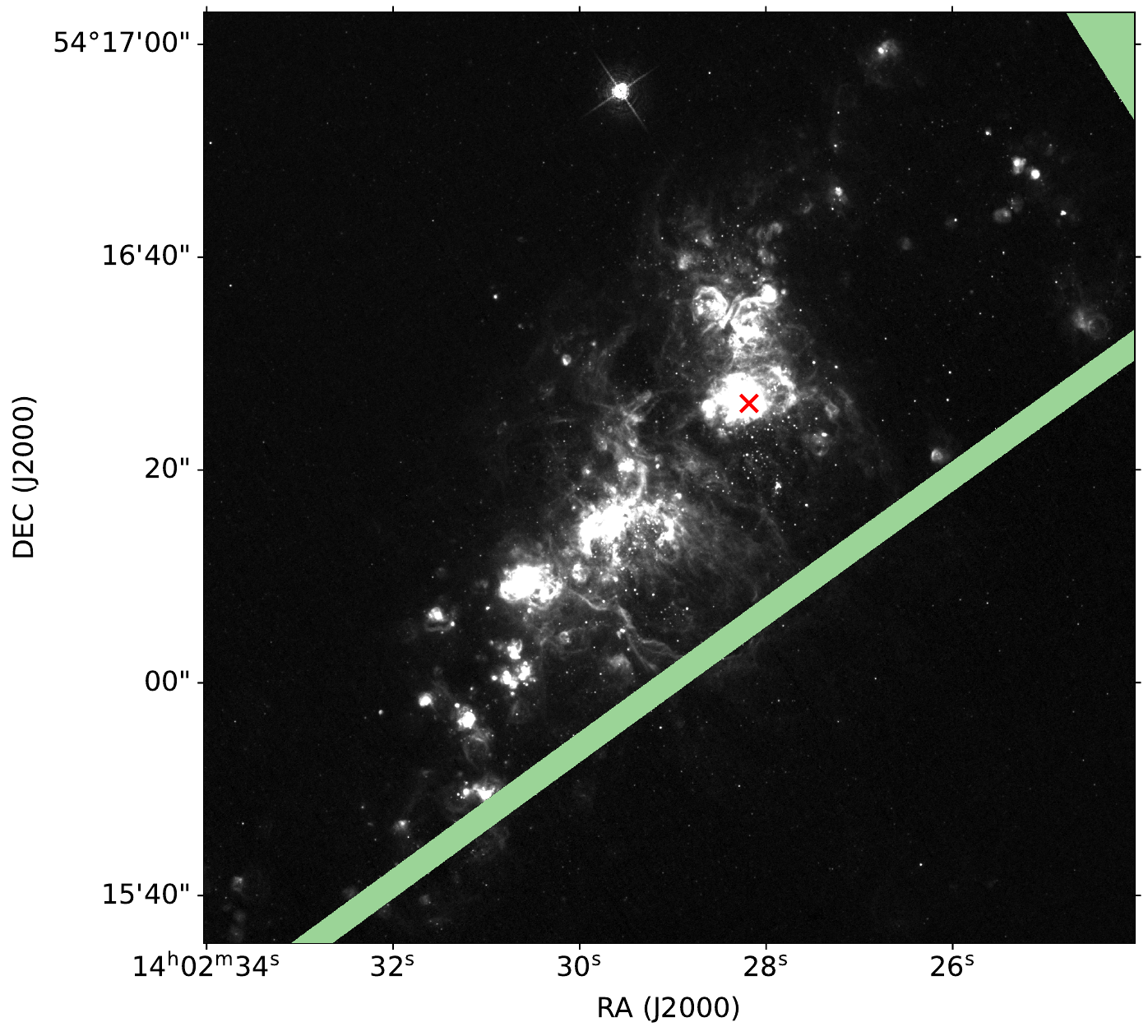}}
	\caption{ NGC 5447 in H$\alpha$+continuum from the Hubble Legacy Archive\protect\footnotemark. The pale green area was not observed. The red x marks the centre of the area in which we measured the spectrum. }
	\label{fig:Halpha_N4557}
\end{figure}

NGC\,5447 is the \hii region in which we do not see a flattening at frequencies down to 54\,MHz (top middle panel of Fig.~\ref{fig:SEDs}). There are no known background radio sources that could affect the flux density in the location of NGC\,5447. If we examine the spectral index of NGC\,5447 (top middle panel of Fig.~\ref{fig:zoom_ins}), we see that around the centre of the region the spectral index is not inverted. There are only two patches with an inverted spectrum near the edges of the \hii region. It is probably the case that the area in which we measure the spectrum, does not include the parts of the region where there is a flattening. If we look at the H$\alpha$+continuum map of NGC\,5447 (Fig.~\ref{fig:Halpha_N4557}), it is revealed that this region consists of multiple distinct condensations, which can also be classified as separate \hii regions. The area where we measure the spectrum in previous section is centred on the north-west condensation. We also try centring the area on the south-east condensation, but this does not change the result and we still do not detect a low frequency flattening. Our result can be explained if the free-free absorption only occurs in small condensations and our resolution is not sufficient to separate the emission with the turnover from the surroundings. This idea could be tested by higher resolution imaging using the international LOFAR stations. This region is also an exception in the work by \cite{Israel_1975} because it is the only region in which they did not see a core component in 6\,cm observations. With the perspective we get from our results, that could also be a consequence of the radio emission being produced by multiple distinct condensations and not one core.

Alternatively, different properties of NGC\,5447 could be because of its position within M\,101. It is located next to the big hole in the disk of M101 that hosts only hot X-ray gas and potentially strong vertical magnetic fields, which could be a result of heating by magnetic reconnection \citep{Wezgowiec_2022}. It is possible that the hole influenced the structure of NGC\,5447. In addition to that, the \hii regions on the East of the galaxy could be different from NGC\,5447 because of the tidal interaction of M101 with the galaxy to the east that also distorts the magnetic arm of M\,101 \citep{Wezgowiec_2022}.

\footnotetext{Based on observations made with the NASA/ESA Hubble Space Telescope, and obtained from the Hubble Legacy Archive, which is a collaboration between the Space Telescope Science Institute (STScI/NASA), the Space Telescope European Coordinating Facility (ST-ECF/ESA) and the Canadian Astronomy Data Centre (CADC/NRC/CSA).}

\subsection{Implication for spectral flattening and radio--FIR correlation}
\label{ss:radio_fir_corelation}

The nearly linear, tight correlation of far-infrared (FIR), and radio luminosity of star-forming galaxies is an important property to measure extinction-free star formation rates \citep{Yun_2001}. This correlation can be explained if the cosmic ray electrons do not escape the galaxy in the so-called calorimetric model \citep{Voelk_1989}. However, in such a case the radio continuum spectra should be steep with integrated radio spectral indices around $\alpha\approx-1.1$ \citep[see e.g.][]{Lacki_2010}. The integrated radio spectral index of galaxies is an important proxy for energy losses and transport effects of cosmic-ray electrons. Synchrotron and inverse Compton radiation losses tend to steepen the radio continuum spectrum, whereas Bremsstrahlung and ionisation losses tend to flatten the spectrum. With our new results we can now quantify the importance of other effects, namely of free--free absorption. The mean emission measure of our H\,{\sc ii} regions is $(7000\pm2600)\,\rm pc\,cm^{-6}$ for the internal model. According to Equation\,\eqref{equation_opacity} the optical depth is $(2.3\pm0.8)\cdot10^{-3}$ at 1\,GHz and $0.28\pm0.11$ at 100\,MHz. This means that the decrement due to free--free absorption can be neglected at 1\,GHz and is of order 30\% at 100\,MHz. We expect that integrated over the entire galaxy, these values are even smaller as the free--free absorption occurs only in regions with dense ionised gas. This means that in regular star forming galaxies, free--free absorption as a cause for spectral flattening may be neglected at frequencies $\gtrsim$100\,MHz. Only in the nuclei of starburst galaxies may such an explanation be worthwhile as both observations \citep{Adebahr_2013} and simulations \citep{Werhahn_2021a} suggest.

The observed flat radio spectral indices (Fig.~\ref{fig:SEDs}, Table \ref{tab:fit_res}), flatter than the canonical injection spectral indices, hint at other energy losses at work. Because they are not strongly frequency-dependent, we do not see a strong downturn at low frequencies. Also, the contribution of free--free emission will flatten the spectrum further at higher frequencies. A positive correlation between spectral curvature and neutral gas density was already found by \citet{Gajovic_2024} in M\,51 corroborating the importance of low energy ionisation losses of the cosmic-ray electrons.

\section{Conclusions}
\label{section_conclusions}

We present new radio continuum maps of M\,101 at 54 and 1370\,MHz. We study the giant \hii regions at low radio frequencies. The centre of the galaxy and five giant \hii regions (NGC 5447, NGC 5455, NGC 5461, NGC 5462, and NGC 5471) are selected for the analysis because they are bright at 144\,MHz.

The spectral index between 54 and 144 MHz is negative in the very centres of the \hii regions. The radio spectra of five out of six giant \hii regions in M\,101 show a flattening below 100\,MHz. We do not see a turnover in the spectrum of NGC\,5447, likely because it consist of several smaller condensations. The flattening observed in the majority of the selected regions can be well fitted with both the internal and external free-free absorption model and our data do not show preference for either model. The resulting EMs are about a factor of two lower for the external model and that is more in line with the EMs estimated from H$\alpha$ fluxes. This is in indication that the external free-free absorption model is better for \hii regions, but because of high uncertainties it is only an indication.

Low-frequency flattening is only present in the very centre of the giant \hii regions. The area where radio spectrum is measured has to be smaller than 1.5 kpc (corresponds to 45\arcsec\ in M\,101) to detect a flattening. In the future, to detect the low frequency flattening in more distant \hii regions, high resolution observations will have to be preformed with the SKA and LOFAR international stations.

Our findings indicate that free--free absorption at frequencies around 100\,MHz has an optical depth of $\approx$0.3 (Sect.\,\ref{ss:radio_fir_corelation}). This means free--free absorption has to be taken into account when observing \hii regions. Globally, however, the effect of free--free absorption is much smaller. This has important consequences when using radio continuum observations to measure extinction-free star-formation rates on kiloparsec scales at frequencies of $\lesssim$100\,MHz \citep[e.g.][]{Heesen_2019}.

\acknowledgements{ We thank the anonymous referee for their comments, which helped to strengthen the conclusions of the paper. We thank M. Weżgowiec for providing the {\sc fits} files for the 4850\,MHz maps.
LG and MB acknowledge funding by the Deutsche Forschungsgemeinschaft (DFG, German Research Foundation) under Germany's Excellence Strategy -- EXC 2121 ``Quantum Universe'' -- 390833306. 
BA and DJB acknowledge funding from the German Science Foundation DFG, via the Collaborative Research Center SFB1491 ‘Cosmic Interacting Matters -- From Source to Signal’.
FdG acknowledges support from the ERC Consolidator Grant ULU 101086378. 
LOFAR \citep{vanHarleem_2013} is the Low Frequency Array designed and constructed by ASTRON. It has observing, data processing, and data storage facilities in several countries, which are owned by various parties (each with their own funding sources), and that are collectively operated by the ILT foundation under a joint scientific policy. The ILT resources have benefited from the following recent major funding sources: CNRS-INSU, Observatoire de Paris and Université d'Orl\'eans, France; BMBF, MIWF-NRW, MPG, Germany; Science Foundation Ireland (SFI), Department of Business, Enterprise and Innovation (DBEI), Ireland; NWO, The Netherlands; The Science and Technology Facilities Council, UK; Ministry of Science and Higher Education, Poland; The Istituto Nazionale di Astrofisica (INAF), Italy. This research made use of the Dutch national e-infrastructure with support of the SURF Cooperative (e-infra 180169) and the LOFAR e-infra group. The J\"ulich LOFAR Long Term Archive and the German LOFAR network are both coordinated and operated by the J\"ulich Supercomputing Centre (JSC), and computing resources on the supercomputer JUWELS at JSC were provided by the Gauss Centre for Supercomputing e.V. (grant CHTB00) through the John von Neumann Institute for Computing (NIC). This research made use of the University of Hertfordshire high-performance computing facility and the LOFAR-UK computing facility located at the University of Hertfordshire and supported by STFC [ST/P000096/1], and of the Italian LOFAR IT computing infrastructure supported and operated by INAF, and by the Physics Department of Turin university (under an agreement with Consorzio Interuniversitario per la Fisica Spaziale) at the C3S Supercomputing Centre, Italy. The research leading to these results has received funding from the European Research Council under the European Union's Seventh Framework Programme (FP/2007-2013) / ERC Advanced Grant RADIOLIFE-320745. This research made use of the Python Kapteyn Package \citep{KapteynPackage}. Partly based on observations made with the NASA/ESA Hubble Space Telescope, and obtained from the Hubble Legacy Archive, which is a collaboration between the Space Telescope Science Institute (STScI/NASA), the Space Telescope European Coordinating Facility (ST-ECF/ESA) and the Canadian Astronomy Data Centre (CADC/NRC/CSA).}

\bibliography{bibtex}{}
\bibliographystyle{aa}

\newpage
\begin{appendix}

\onecolumn 

\section{Mask for the integrated spectrum}
\label{a:mask_int}
\vspace{-0.5cm}
\begin{figure*}[h!]
    \centering
	\resizebox{0.9\hsize}{!}{\includegraphics{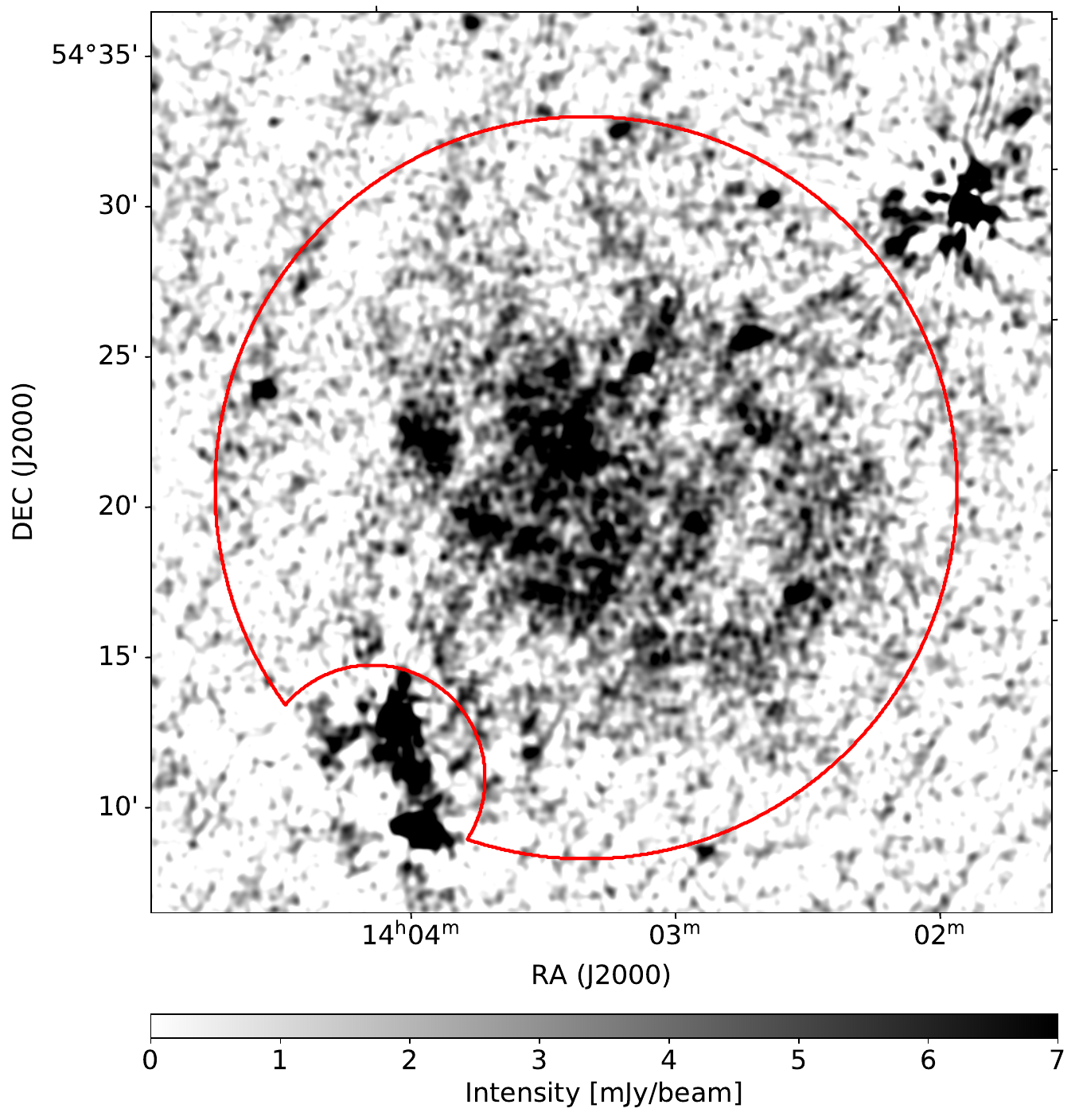}}
	\caption{The region in which we calculated the integrated spectrum of M101. The region is plotted in red on top of the 54\,MHz LBA map. The artefacts around background sources are most prominent in the LBA map so we use that map to make sure the artefacts are excluded.}
\end{figure*}

\newpage
\section{Archival integrated flux density measurements for M 101}
\label{a:integrated}

\begin{table*}[h!]
        \caption{Integrated flux density of M\,101 from the literature and our observations. }
        \label{tab:integrated}
        \centering{
        \newcolumntype{0}{>{\centering\arraybackslash} m{1.8cm} }
        \newcolumntype{1}{>{\centering\arraybackslash} m{2.8cm} }
        \newcolumntype{2}{>{\centering\arraybackslash} m{1.9cm} }
        \newcolumntype{3}{>{\centering\arraybackslash} m{1.8cm} }
        \newcolumntype{4}{>{\centering\arraybackslash} m{2.2cm} }
        \newcolumntype{5}{>{\centering\arraybackslash} m{1.8cm} }
        \renewcommand{\arraystretch}{1.5}
        \begin{tabular}{@{} 0 1 2 3 4 5 @{}}
                \toprule
                \toprule
                Frequency [MHz] & Flux density [Jy] & Instrument type & Background sources removed & Beam size & Reference \\
                \midrule
        54\tablefootmark{a} & $5.02\pm 0.51$ & Interferometer & + & $26\farcs5 \times 20\arcsec$ & 1  \\
        57 & $6.5\pm 2.5$ & Interferometer & - &   $\ \ 7\arcmin \times 6\arcmin$ & 2  \\
        144\tablefootmark{a} & $3.18\pm 0.32$ & Interferometer & + & $26\farcs5 \times 20\arcsec$ &  3  \\
        178 & $2.4\pm 0.5$ & Interferometer & - & $\ 23\arcmin \times 18\arcmin$ & 4  \\
        610 & $1.4\pm 0.1$ & Interferometer & + & $\ \ \ \ \ \ 1\arcmin \times 1\arcmin12\arcsec$ & 5  \\ 
        750 & $1.3\pm 0.3$ & Single dish & - & $18\arcmin30\arcsec \times 18\arcmin30\arcsec$ & 6  \\
        920 & $1.3\pm 0.2$ & Transit & - & $\sim 8\arcmin$ in E-W direction & 7  \\
        1370\tablefootmark{a}  & $0.537\pm 0.027$ & Interferometer & + &$15\farcs 1\times 11\farcs 0$ & 1  \\
        1400 & $0.75\pm 0.04$ & Interferometer & - & $45\arcsec \times 45\arcsec$ & 8  \\  
        1400 & $0.81\pm 0.10$ & Single dish & - & $11\arcmin40\arcsec \times 11\arcmin40\arcsec $ & 9  \\  
        1400 & $0.8\pm 0.1$ & Interferometer & - &  $4\arcmin\times 4\arcmin$  & 10  \\
        1410 & $0.82\pm 0.21$ & Single dish & - & $10\arcmin \times 10\arcmin$ & 6  \\
        1415 & $0.85\pm 0.13$ & Interferometer & - & $4\arcmin\times 30\arcmin$ & 11  \\
        2700 & $0.52\pm 0.06$ & Single dish & + & $\sim\! 4\arcmin30\arcsec\times 4\arcmin30\arcsec\ \ \ $ & 12  \\
        2700 & $0.48\pm 0.03$ & Single dish & + & $4\arcmin24\arcsec\times 4\arcmin24\arcsec$ & 13  \\
        4750 & $0.335\pm 0.020$ & Single dish & + & $2\arcmin27\arcsec\times 2\arcmin27\arcsec$ & 12  \\
        4850 & $0.31\pm 0.02$ & Single dish & + & $2\arcmin30\arcsec\times 2\arcmin30\arcsec$ & 13  \\
        4850\tablefootmark{a} & $0.308\pm 0.015$ & Combined & + & $30\arcsec\times 30\arcsec$ & 14  \\
        10700 & $0.207\pm 0.020$ & Single dish & + & $1\arcmin30\arcsec\times 1\arcmin30\arcsec$ & 12  \\

        \bottomrule
        \end{tabular}}
        \tablebib{(1) This paper; (2) \citet{Israel_1990}; (3) \cite{Shimwell_2022}; (4) \cite{Caswell_1967}; (5) \cite{Israel_1975}; (6) \cite{deJong_1965}; (7) \cite{KurilChik_1966}; (8) \cite{Condon_2002}; (9) \cite{White_1992}; (10) \cite{Rogstad_1971}; (11) \cite{deLaBeaujardiere_1968}; (12) \cite{Graeve_1990}; (13) \cite{Berkhuijsen_2016};   (14) \cite{Wezgowiec_2022}; 
        \\}
        \raggedright{\tablefoottext{a}{Map used in this work.}}
        
\end{table*}

\newpage

\section{Spectral index error map}
\label{a:error_map}

\begin{figure*}[h!]
    \centering
	\resizebox{0.9\hsize}{!}{\includegraphics{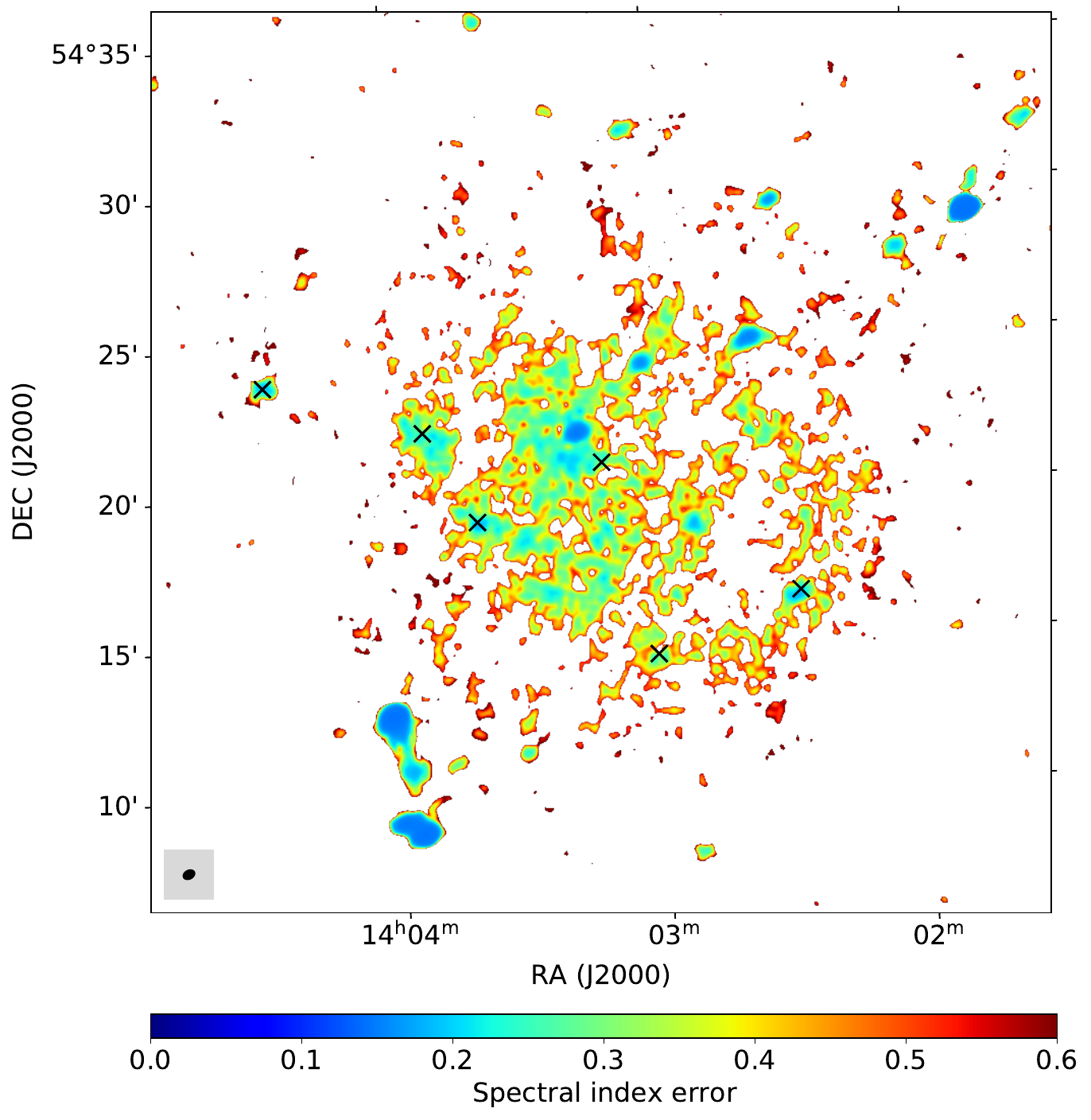}}
	\caption{Spectral index error map between 54\,MHz and 144\,MHz. White areas below below 2$\sigma$ are excluded. The beam size is shown with the black ellipse in the bottom left corner. The centres of the studied \hii regions are marked with an x to make them easier to find.}
	\label{fig:spix_err}
\end{figure*}

\end{appendix}
\end{document}